\documentclass{article}
\usepackage[margin=1in]{geometry}
\usepackage{graphicx}
\usepackage{mathptmx}
\usepackage{amsmath}
\usepackage{booktabs}
\usepackage{doi}
\usepackage{natbib}
\usepackage{xcolor} 
\usepackage{hyperref} 
\hypersetup{
  colorlinks = true,
  linkcolor = blue,
  citecolor = blue,
  urlcolor = blue
} 
\usepackage[utf8]{inputenc} 

\begin{document}

\title{Modelling rankings in R: the \textbf{PlackettLuce}
  package\thanks{This research is part of cooperative agreement 
  AID-OAA-F-14-00035, which was made possible by the generous support 
  of the American people through the United States Agency for 
  International Development (USAID). This work was implemented as 
  part of the CGIAR Research Program on Climate Change, Agriculture 
  and Food Security (CCAFS), which is carried out with support from 
  CGIAR Fund Donors and through bilateral funding agreements. For 
  details please visit https://ccafs.cgiar.org/donors. The views 
  expressed in this document cannot be taken to reflect the official 
  opinions of these organisations.
  The work of David Firth and Ioannis Kosmidis was
  supported by the Alan Turing Institute under EPSRC grant
  EP/N510129/1, and part of it was completed when the Ioannis
  Kosmidis was a Senior Lecturer at University College London.}}

\author{Heather L. Turner\thanks{
           Department of Statistics, University of Warwick, UK.
           ORCiD: 0000-0002-1256-3375.
           \href{mailto:ht@heatherturner.net}{ht@heatherturner.net}}
           \and
           Jacob van Etten\thanks{
           Bioversity International, Costa Rica.
           ORCiD: 0000-0001-7554-2558.}
           \and
           David Firth\thanks{
           Department of Statistics, University of Warwick, UK; and
           The Alan Turing Institute, London, UK.
           ORCiD: 0000-0003-0302-2312.}
           \and
           Ioannis Kosmidis\thanks{
           Department of Statistics, University of Warwick, UK; and
           The Alan Turing Institute, London, UK.
           ORCiD: 0000-0003-1556-0302.}
           }
           
\maketitle

\begin{abstract}
  This paper presents the R package \textbf{PlackettLuce}, which implements a
  generalization of the Plackett-Luce model for rankings data. The
  generalization accommodates both ties (of arbitrary order) and partial rankings
  (complete rankings of subsets of items). By default, the implementation adds a set of
  pseudo-comparisons with a hypothetical item, ensuring that the underlying 
  network of wins and losses between items is always strongly connected. In this way, the 
  worth of each item always has a finite maximum likelihood estimate, with finite
  standard error. The use of pseudo-comparisons also has
  a regularization effect, shrinking the estimated parameters towards equal
  item worth. In addition to standard methods for model summary,
  \textbf{PlackettLuce} provides a method to compute quasi standard errors for
  the item parameters. This provides the basis for comparison intervals that 
  do not change with the choice of identifiability constraint placed on the 
  item parameters. Finally, the package provides a method for model-based
  partitioning using covariates whose values vary between rankings, enabling the
  identification of subgroups of judges or settings with different item 
  worths. The features of the package are demonstrated through application to 
  classic and novel data sets.
\end{abstract}

\hypertarget{introduction}{%
\section{Introduction}\label{introduction}}

Rankings data arise in a range of applications, such as sport tournaments and
consumer studies. In rankings data, each observation is an ordering of a set
of items. A classic model for such data is the Plackett-Luce model, which
stems from Luce's axiom of choice \citep{Luce1959, Luce1977}. This axiom
states that the odds of choosing one item, \(i_1\), over another, \(i_2\), do not
depend on the other items available for selection in that choice.

Suppose we have a set of \(J\) items
\[S = \{i_1, i_2, \ldots, i_J\}.\]
Under Luce's axiom, the probability of selecting some item \(i_j\)
from \(S\) is given by
\[P(i_j | S) = \frac{\alpha_{i_j}}{\sum_{i \in S} \alpha_i},\]
where \(\alpha_i \ge 0\) represents the \textbf{worth} of item \(i\). The worth is a
latent characteristic, with higher values indicating that an item is more
likely to be selected. In practice, it could represent the ability
of an athlete or the attractiveness of a product, for example.

A ranking of \(J\)
items can be viewed as a sequence of independent choices: first choosing the
top-ranked item from all items, then choosing the second-ranked item from the
remaining items and so on. Let \(a \succ b\) denote that \(a\) is ranked higher
than \(b\). It follows from Luce's axiom that the probability of the ranking
\({i_1 \succ \ldots \succ i_J}\) is
\begin{equation}
\prod_{j=1}^J \frac{\alpha_{i_j}}{\sum_{i \in A_j} \alpha_i},
\label{eq:likelihood}
\end{equation}
where \(A_j\) is the set of alternatives \(\{i_j, i_{j + 1}, \ldots, i_J\}\) from
which item \(i_j\) is chosen. The above model is also derived in \citet{Plackett1975},
hence the name Plackett-Luce model.

In this paper, we present the R package \textbf{PlackettLuce} \citep{Turner2019a}.
This package implements an extension of the Plackett-Luce model that allows for
ties in the rankings. The model can be applied to either complete or partial
rankings (complete rankings of subsets of items).
\textbf{PlackettLuce} offers a choice of algorithms to fit the model via
maximum likelihood. Pseudo-rankings, i.e.~pairwise comparisons with a
hypothetical item, are used to ensure that the item worths always have finite
maximum likelihood estimates (MLEs) with finite standard error. Methods are provided to obtain
different parameterizations with corresponding standard errors or
quasi-standard errors (that do not change with the identifiability constraint).
There is also a method to fit Plackett-Luce trees, which partition the rankings
by covariate values to identify subgroups with distinct Plackett-Luce models.

In the next section, we review the available software for modelling rankings and
make comparisons with \textbf{PlackettLuce}. Then in Section
\ref{plackett-luce-model-with-ties} we describe the Plackett-Luce model with
ties and the methods implemented in the package for model-fitting and inference.
Plackett-Luce trees are introduced in Section
\ref{plackett-luce-trees}, before we conclude the paper with a discussion in
Section~\ref{discussion}.

\hypertarget{software-for-modelling-rankings}{%
\section{Software for modelling rankings}\label{software-for-modelling-rankings}}

First, in Section~\ref{fitting-the-plackett-luce-model}, we consider
software to fit the standard Plackett-Luce model. Then in
Section~\ref{features-beyond-fitting-the-standard-model}, we review the
key packages in terms of their support for features beyond fitting the
standard model. Finally in Section~\ref{other-software-for-modelling-rankings}
we review software implementing alternative models for rankings data and
discuss how these approaches compare with the Plackett-Luce model.

\hypertarget{fitting-the-plackett-luce-model}{%
\subsection{Fitting the Plackett-Luce model}\label{fitting-the-plackett-luce-model}}

Sections \ref{fitting-as-a-log-linear-model} and
\ref{fitting-as-a-cox-proportional-hazards-model} describe two ways the
Plackett-Luce model can be fitted with standard statistical software. Section
\ref{specialized-software} reviews packages that have been specifically
developed to fit the Plackett-Luce model or a related genralization. The
performance of the key packages for R are compared in Section
\ref{performance-of-software-to-fit-the-plackett-luce-model}.

\hypertarget{fitting-as-a-log-linear-model}{%
\subsubsection{Fitting as a log-linear model}\label{fitting-as-a-log-linear-model}}

Each choice in the ranking can be considered as a multinomial observation,
with one item observed out of a possible set. The ``Poisson trick''
\citep[see, for example,][]{Baker1994} expresses these multinomial observations as
counts (one or zero) of each possible outcome within each choice. For example,
the ranking \{item 3\} \(\succ\) \{item 1\} \(\succ\) \{item 2\} is equivalent to two
multinomial observations: \{item 3\} from \{item 1, item 2, item 3\} and \{item 1\}
from \{item 2, item 3\}. These observations are expressed as counts as follows:

\begin{verbatim}
     choice item 1 item 2 item 3 count
[1,]      1      1      0      0     0
[2,]      1      0      1      0     0
[3,]      1      0      0      1     1
[4,]      2      1      0      0     1
[5,]      2      0      1      0     0
\end{verbatim}

The Plackett-Luce model is then equivalent to a log-linear model of \texttt{count},
with \texttt{choice} included as a nuisance factor to ensure the
multinomial probabilities sum to one, and the dummy variables for each item
(\texttt{item\ 1}, \texttt{item\ 2}, \texttt{item\ 3})
included to estimate the item worth parameters.

As a result, the Plackett-Luce model can be fitted using standard software for generalized
linear models (GLMs). However, expanding the rankings to counts of each possible
choice is a non-standard task. Also, the choice factor will have many levels:
greater than the number of rankings, for rankings of more than two objects.
A standard GLM fitting function such as \texttt{glm} in \texttt{R} will be slow to fit the model,
or may even fail as the corresponding model matrix will be too large to fit
in memory. This issue can be circumvented by ``eliminating'' the nuisance \texttt{choice} factor, by
taking advantage of the diagonal submatrix that would appear in the Fisher
Information matrix to compute updates without requiring the full design matrix.
This feature is available, for example, in the \texttt{gnm} function from the \texttt{R}
package \textbf{gnm} \citep{Turner2018b}. Even then, the model-fitting may be relatively
slow, given the expansion in the number of observations when rankings are
converted to counts. This issue is exacerbated when ties are allowed, since the
number of possible outcomes increases quickly with the number of items.

\hypertarget{fitting-as-a-cox-proportional-hazards-model}{%
\subsubsection{Fitting as a Cox proportional hazards model}\label{fitting-as-a-cox-proportional-hazards-model}}

For complete rankings, the probability in Equation~\eqref{eq:likelihood}
is equivalent to that for a special case of
the \emph{rank-ordered logit model} or \emph{exploded logit model} \citep{Allison1994}. This model
assumes a certain utility \(U_{ri}\) for each item \(i\) and respondent
\(r\), which is modelled as
\[U_{ri} = \mu_{ri} + \epsilon_{ri}.\]
Here \(\mu_{ri}\) may be decomposed into a linear function of
explanatory variables, while the \(\{\epsilon_{ri}\}\) are independent and
identically distributed with an extreme value distribution. The
standard Plackett-Luce model is then given by a rank-ordered logit
model with \(U_{ri} = \log(\alpha_i) + \epsilon_{ri}\).

\citet{Allison1994} demonstrated that the rank-ordered logit model can be estimated by
fitting a Cox proportional hazards model with equivalent partial likelihood.
Therefore, specialist software is not required to fit the model. Software
such as the \texttt{rologit} function in \texttt{Stata} or the \texttt{rologit} function from the
\texttt{R} package \textbf{ROlogit} \citep{Tan2018} provide wrappers to methods for the Cox
proportional hazards model, so that users do not have to specify the equivalent
survival model.

Compared to the log-linear approach, the rank-ordered logit model has
the advantage that the ranks can be modelled directly without expanding into
counts. Nonetheless, the model fitting requires
a dummy variable for each item and a stratification factor for each unique
ranking. Therefore, the estimation becomes slow with a high number of items
and/or unique rankings.

\hypertarget{specialized-software}{%
\subsubsection{Specialized software}\label{specialized-software}}

Several software packages have been developed to model rankings data,
which provide specialized functions to fit the Plackett-Luce model. Rather than
re-expressing the model so that it can be estimated with a standard modelling
function, these specialized functions take a direct approach to fitting the
Plackett-Luce model.

One such approach is Hunter's \citeyearpar{Hunter2004}
minorization-maximization (MM) algorithm to maximize the likelihood of
the observed data under the Plackett-Luce model.
This algorithm is used by the R package \textbf{StatRank} and
is one of the algorithms available in the Python package \textbf{choix} \citep{Maystre2018}.
Alternatively the likelihood can be maximised directly using a generic
optimisation method such as the
Broyden--Fletcher--Goldfarb--Shanno (BFGS) algorithm. This is another
option offered by \textbf{choix} and is the method used in the
R packages \textbf{pmr} \citep{Lee2015} and \textbf{hyper2} \citep{Hankin2019}. A third
option offered by \textbf{choix} is the `Luce spectral algorithm', which in a
single iteration (LSR) gives a close approximation to the
MLEs and may be used iteratively (I-LSR)
to give a maximum likelihood solution \citep{Maystre2015}.

In addition, a number of algorithms have been proposed to fit
Plackett-Luce mixture models. These models assume the rankings are observations
from \(G\) subgroups, each with a distinct Plackett-Luce model. The group
membership probabilities are estimated as part of the model-fitting. Therefore, software to
fit Plackett-Luce mixture models can be used fit the standard Plackett-Luce
model by fixing the number of groups, \(G\), to be one. The \texttt{R} package
\textbf{mixedMem} \citep{Wang2015} implements a variational Expectation-Maximization
(EM) algorithm, which returns MLEs
in the case of a single group. The \texttt{R} package \textbf{PLMIX}
\citep{Mollica2014, Mollica2017} uses a Bayesian framework and offers the choice to
maximize the posterior distribution via an EM algorithm, which reduces
to the MM algorithm with noninformative hyperparameters, or to
simulate the posterior distribution using Markov-chain Monte-Carlo
(MCMC) techniques.

\textbf{PlackettLuce} offers both iterative scaling,
which is equivalent to the MM algorithm for the standard Plackett-Luce
model, and generic optimization using either BFGS or a limited memory
variant (L-BFGS) via the \texttt{R} package \textbf{lbfgs} package \citep{Coppola2014}.

\hypertarget{performance-of-software-to-fit-the-plackett-luce-model}{%
\subsubsection{Performance of software to fit the Plackett-Luce model}\label{performance-of-software-to-fit-the-plackett-luce-model}}

Even using specialized software, the time taken to fit a Plackett-Luce model
can scale poorly with the number of items and/or the number of rankings.
Table \ref{tab:timings-kable} shows some example run times for fitting the
Plackett-Luce model with specialized software available in R,
including \textbf{ROlogit} for comparison. The run times are
presented for three different data sets taken from the PrefLib library of
preference data \citep{Mattei2013}. The data sets are selected to
represent a range of features, as summarised in Table
\ref{tab:data-features}.

As expected, \textbf{ROlogit}, based on \texttt{R}'s \texttt{coxph} function, shows a marked
drop in performance when the number of unique rankings is high. A similar
drop off is seen for \textbf{hyper2} where the time is
spent specifying the likelihood function for each unique ranking.
Although \textbf{StatRank} uses an MM algorithm, the same as used by
\textbf{PlackettLuce} and \textbf{PLMIX} in these benchmarks, it shows poor performance for
a moderate number of items (\(\ge\) 10) and/or a high number of unique rankings.
Similarly \textbf{pmr} which uses the same optimization method as
\textbf{hyper2} becomes impractical to use with a moderate number of items.

Clearly, the details of the implementation are as important as the
estimation method used. \textbf{PlackettLuce} copes well with these
moderately-sized data sets and is comparable to \textbf{mixedMem}.
\textbf{PlackettLuce} is faster than \textbf{mixedMem} on the Netflix data as it can
work with aggregated unique rankings. \textbf{PLMIX} is consistently
fast; the superior performance over \textbf{PlackettLuce} when both the
number of items and the number of unique rankings is high can be explained by
its core functionality being implemented in \texttt{C++} rather than pure \texttt{R} code.

\begin{table}[t]

\caption{\label{tab:data-features}Features of example data sets from PrefLib \citep{Mattei2013}. The Netflix data were collated by \citep{Bennett2007}.}
\centering
\begin{tabular}{lrrr}
\toprule
  & Rankings & Unique rankings & Items\\
\midrule
Netflix & 1256 & 24 & 4\\
T-shirt & 30 & 30 & 11\\
Sushi & 5000 & 4926 & 10\\
\bottomrule
\end{tabular}
\end{table}

\begin{table}[t]

\caption{\label{tab:timings-kable}Run times for fitting the Plackett-Luce model to data sets summarised in
Table \ref{tab:data-features} using different R packages. See Appendix \ref{app:run-times} for details and code.}
\centering
\begin{tabular}{lrrrrrrr}
\toprule
\multicolumn{1}{c}{ } & \multicolumn{7}{c}{Time elapsed (s)} \\
\cmidrule(l{3pt}r{3pt}){2-8}
  & PlackettLuce & hyper2 & PLMIX & pmr & StatRank & ROlogit & mixedMem\\
\midrule
Netflix & 0.042 & 0.115 & 0.535 & 0.586 & 0.648 & 1.462 & 0.388\\
T-shirt & 0.023 & 0.123 & 0.007 & $^{\rm{a}}$ & 7.473 & 0.043 & 0.006\\
Sushi & 1.384 & 83.657 & 0.135 & $^{\rm{a}}$ & 23.074 & 10.991 & 1.879\\
\bottomrule
\multicolumn{8}{l}{\textsuperscript{a} Function fails to complete.}\\
\end{tabular}
\end{table}

\hypertarget{features-beyond-fitting-the-standard-model}{%
\subsection{Features beyond fitting the standard model}\label{features-beyond-fitting-the-standard-model}}

So far, we have compared packages in terms of their ability to fit the standard
Plackett-Luce model. In practice, rankings data can have features that require
a generalization of the standard model, or that can cause issues with maximum
likelihood estimation. In addition, not all packages provide facilities for
inference on the estimated item worths.

In this section, we compare the key packages for fitting the Plackett-Luce model
in terms of their ability to meet these additional requirements.

\hypertarget{partial-rankings}{%
\subsubsection{Partial rankings}\label{partial-rankings}}

As the number of items increases, it is typically more common to
observe partial rankings than complete rankings. Partial rankings can be of two
types: \emph{subset rankings}, where subsets of items are completely ranked in each
observation, and \emph{top \(n\) rankings}, where the top \(n\) items are selected and
the remaining items are unranked, but implicitly ranked lower than the top \(n\).
Both types can be accommodated by the Plackett-Luce model but imply a different
form for the likelihood. For subset rankings, the set of alternatives in the
denominator of \eqref{eq:likelihood} will only comprise remaining items from
the subset of items in the current ranking. For top \(n\) rankings the set of
alternatives is always the full set of items minus the items that have been
selected so far in the current ranking.

Given this difference in the likelihood, implementations tend to support only
one type of partial ranking. \textbf{PlackettLuce} and \textbf{choix} handle
subset rankings only, while \textbf{PLMIX}, \textbf{ROlogit} and \textbf{mixedMem} handle
top \(n\) rankings only. The documentation for \textbf{StatRank} suggests that it
supports subset rankings, but we were unable to validate such support. \textbf{hyper2}
can handle both types of partial ranking, since it requires the likelihood to be
specified for each observation. However, \textbf{hyper2} provides a utility function
for subset rankings making these easier to fit with the package.

Table \ref{tab:nascar} shows run times for fitting the Plackett-Luce model to
the NASCAR subset rankings from \citet{Hunter2004}. These results illustrate that
\textbf{PlackettLuce} is more efficient than \textbf{hyper2} for modelling subset
rankings of a relatively large number of items.

\begin{table}[t]

\caption{\label{tab:nascar}Run times for fitting the Plackett-Luce model to the NASCAR data from \citet{Hunter2004}. All rankings are unique.}
\centering
\begin{tabular}{rrrrr}
\toprule
\multicolumn{3}{c}{Features of NASCAR data} & \multicolumn{2}{c}{Time elapsed (s)} \\
\cmidrule(l{3pt}r{3pt}){1-3} \cmidrule(l{3pt}r{3pt}){4-5}
Rankings & Items & Items per ranking & PlackettLuce & hyper2\\
\midrule
36 & 83 & 42-43 & 0.199 & 36.123\\
\bottomrule
\end{tabular}
\end{table}

\hypertarget{tied-ranks}{%
\subsubsection{Tied ranks}\label{tied-ranks}}

\textbf{ROlogit} can handle tied ranks as tied survival times in the
underlying Cox proportional hazards model. The ties are
accommodated in the likelihood by summing over all possible strict rankings
that could have resulted in the observed tie. This assumes that one of these
strict rankings is correct, but a lack of precision means we observe a tie.

However in some contexts, a tie is a valid outcome, for example, an
equal number of goals in a sports match. As far as we are aware,
\textbf{PlackettLuce} is the only software to model ties as an explicit outcome
based on item worth, through an extension of the Plackett-Luce model
described in Section~\ref{plackett-luce-model-with-ties}.

\hypertarget{inference}{%
\subsubsection{Inference}\label{inference}}

In order to conduct frequentist inference on the item parameters, it is
necessary to compute standard errors. The only packages based
on maximum likelihood estimation with this functionality are \textbf{PlackettLuce}
and \textbf{ROlogit}. The computation of the variance-covariance
matrix in \textbf{PlackettLuce} is currently based on the methodology for
generalized linear models, so the rankings are internally expanded as counts as
described in Section~\ref{fitting-as-a-log-linear-model}.

\textbf{PLMIX} allows for a fully Bayesian inference based on the approximation
of the posterior distribution.

\hypertarget{disconnected-or-weakly-connected-networks}{%
\subsubsection{Disconnected or weakly connected networks}\label{disconnected-or-weakly-connected-networks}}

In some cases, the MLE does not exist, or has
infinite standard error. This happens when the network of
wins and losses implied by the rankings is not strongly connected -- an issue
explored further in Section~\ref{disconnected-or-weakly-connected-networks}.
\textbf{PlackettLuce} ensures that the item worths can always be estimated by
using pseudo-rankings. This approach can be viewed as incorporating a
`shrinkage' prior as in a Bayesian analysis (but with estimation
based on maximization of the posterior density). The \textbf{choix} package
implements methods of regularization that have a similar effect.

Since \textbf{PLMIX} uses a Bayesian framework that incorporates a prior, the
package will always produce finite worth estimates.

\hypertarget{heterogeneity}{%
\subsubsection{Heterogeneity}\label{heterogeneity}}

As noted in Section~\ref{specialized-software}
\textbf{PLMIX} and \textbf{mixedMem} offer the facility to model heterogeneous rankings
via mixture models, which allow for latent groups with different item worth.
This is similar in spirit to the tree models offered by \textbf{PlackettLuce}, which
define such groups through recursive binary splits on attributes of
the judges or other variables that covary with the rankings. This
partitioning approach does not rely on parametric assumptions, whereas the
mixture models assume a heterogeneous sampling distribution.

\hypertarget{summary-of-software-features}{%
\subsubsection{Summary of software features}\label{summary-of-software-features}}

A summary of the features of the main software for Plackett-Luce models is
given in Table \ref{tab:package-summary}. \textbf{pmr} and \textbf{StatRank} are not
included as they do not support any of the features mentioned (with the possible
exception of partial rankings in the case of \textbf{StatRank}, where the documented
support could not be validated).

\begin{table}[t]

\caption{\label{tab:package-summary}Features of key packages for fitting the Plackett-Luce model. ``Inference'' refers to inference beyond parameter estimation. ``Weak network'' is short-hand for a network that is not strongly connected.}
\centering
\begin{tabular}{llllllll}
\toprule
Package & Partial 
 rankings & Ties & Inference & Weak 
 network & Heterogeneity & Covariates & Teams\\
\midrule
PlackettLuce & Subset & Yes & Frequentist & Yes & Tree & No & No\\
ROlogit & Top $n$ & Yes & Frequentist & No & None & Yes & No\\
PLMIX & Top $n$ & No & Bayesian & Yes & Mixture & No & No\\
mixedMem & Top $n$ & No & None & No & Mixture & No & No\\
hyper2 & Any & No & None & No & None & No & Yes\\
choix & Subset & No & None & Yes & None & No & No\\
\bottomrule
\end{tabular}
\end{table}

Some of the software for Plackett-Luce models have features beyond the
scope of \textbf{PlackettLuce} and these features are also included in Table
\ref{tab:package-summary}. \textbf{ROlogit} can model the item
worth as a linear function of covariates. \textbf{hyper2} can handle rankings
of combinations of items, for example, team rankings in sports. This feature is
implemented by extending
the Plackett-Luce model to describe the selection of a
combination \(T_j\) of items from the alternatives, that is
\[\prod_{j=1}^J \frac{\sum_{i \in T_j} \alpha_{i}}{\sum_{i \in A_j} \alpha_i}.\]

\hypertarget{other-software-for-modelling-rankings}{%
\subsection{Other software for modelling rankings}\label{other-software-for-modelling-rankings}}

Although we are focusing on the Plackett-Luce model in this paper, there are
other approaches to modelling rankings data. This section reviews the main
alternatives and the available implementations.

\hypertarget{mallows-bradley-terry-model}{%
\subsubsection{Mallows' Bradley-Terry model}\label{mallows-bradley-terry-model}}

For partial rankings of two items, i.e., paired comparisons, the Plackett-Luce
model reduces to the Bradley-Terry model \citep{Bradley1952}. Mallows' Bradley-Terry
model \citep{Mallows1957} is an alternative way to extend the Bradley-Terry model
for complete rankings. Each ranking is considered as a set of paired comparisons
and their joint probability is modelled assuming independence, normalizing over
all possible rankings. Mallows' Bradley-Terry model is implemented by the
\textbf{eba} and \textbf{prefmod} packages for \texttt{R} \citetext{\citealp{Wickelmaier2004}; \citealp[and][respectively]{Hatzinger2012}}.
Although both packages estimate item worths, neither \textbf{eba}
nor \textbf{prefmod} provide methods for inference based on these estimates.

The \textbf{prefmod} package implements Mallows' Bradley-Terry model as a special
case of a wider class of pattern models, where a pattern is a set of
paired comparisons considered simultaneously. Several extensions of the basic
model are implemented in \textbf{prefmod}: allowing for dependencies
between paired comparisons that involve the same item; allowing for ties and
partial rankings; incorporating item, subject or order effects, and allowing
for latent subject effects via mixture models. However,
not all of these features can be combined. Specifically, item effects,
subject effects due to continuous attributes, order effects and
mixture models are not compatible with partial rankings. Partial
rankings are implemented by treating unranked items as having missing ranks
and all possible rankings including missing values must be considered. This
means that the methods do not scale well to a large number
of items (the package documentation suggests 8 items as a reasonable maximum).

\hypertarget{mallows-distance-based-model}{%
\subsubsection{Mallows distance-based model}\label{mallows-distance-based-model}}

The Mallows distance-based model \citep{Mallows1957} does not model the worth of
items, but rather a global ranking. Specifically, it models the modal ranking
and a dispersion parameter, which describes how quickly the probability of a
ranking decreases with increasing distance from the modal ranking.
Extensions include the generalized Mallows model that allows
a separate dispersion parameter for each free position in the ranking, which
can give some indication of the relative worth. Standard and generalized
versions of Mallows models are implemented in the \textbf{pmr} and \textbf{PerMallows}
\texttt{R} packages \citep{Irurozki2016}.

The \textbf{rankdist} \texttt{R} package implements mixtures of the standard Mallows model for
complete or subset rankings via maximum likelihood \citep{Qian2019}. The
\textbf{BayesMallow} \texttt{R} package \citep{Vitelli2018} implements mixtures of the
standard Mallows model in a Bayesian framework. This has the advantage of
facilitating inference on the global ranking (in each sub-population),
e.g.~95\% highest posterior density intervals on ranking position; probability of an
item appearing in the top 10, etc. In addition, subset rankings,
top \(n\) rankings, missing ranks and ties can all be accommodated in the
Bayesian model.

\hypertarget{insertion-sorting-rank-model}{%
\subsubsection{Insertion Sorting Rank model}\label{insertion-sorting-rank-model}}

The \textbf{Rankcluster} package \citep{Grimonprez2016} implements an alternative
probabilistic model for rankings called the Insertion Sorting Rank model. This
is based on a modal ranking and a parameter for the probability of a ranker
making an implicit paired comparison consistent with the modal ranking.
Heterogeneity is accommodated via a mixture model and the model can be
applied to subset or top-\(n\) rankings. As the package name suggests, the focus
is on clustering subjects by their modal ranking and the probability of
membership, rather than inference on item worths.

\hypertarget{summary}{%
\subsubsection{Summary}\label{summary}}

Table \ref{tab:other-package-summary} provides an overview of software based on
alternative models, compared to \textbf{PlackettLuce}.

\begin{table}[t]

\caption{\label{tab:other-package-summary}Features of key packages using alternative models for rankings, compared to \textbf{PlackettLuce}. Mallows BT = Mallows Bradley-Terry model, Mallows DB = Mallows distance-based model, ISR = Insertion Sorting Rank model. ``Inference'' refers to inference beyond parameter estimation, related to the focus of analysis under ``Focus''.}
\centering
\begin{tabular}{lllllll}
\toprule
Package & Model & Focus & Inference & Partial 
 rankings & Ties & Heterogeneity\\
\midrule
PlackettLuce & PL & Worth & Frequentist & Subset & Yes & Trees\\
eba & Mallows BT & Worth & None & None & No & None\\
prefmod & Mallows BT & Worth & None & Any & Yes & Categorical effects\\
PerMallows & Mallows DB & Ranking & None & None & No & None\\
BayesMallow & Mallows DB & Ranking & Bayesian & Any & Yes & Mixture\\
rankdist & Mallows DB & Ranking & None & Subset & No & Mixture\\
Rankcluster & ISR & Cluster membership & Frequentist & Any & No & Mixture\\
\bottomrule
\end{tabular}
\end{table}

\hypertarget{plackett-luce-model-with-ties}{%
\section{Plackett-Luce model with ties}\label{plackett-luce-model-with-ties}}

In this section, we describe and demonstrate the core features of the
\textbf{PlackettLuce} package: modelling partial rankings with ties,
conducting inference on the worth parameters, and handling ranking networks
that are not strongly connected.

\hypertarget{modelling-partial-rankings-with-ties}{%
\subsection{Modelling partial rankings with ties}\label{modelling-partial-rankings-with-ties}}

The \textbf{PlackettLuce} package permits rankings of the form
\[R = \{C_1, C_2, \ldots, C_J\},\]
where the items in set \(C_1\) are ranked higher than (better than) the items
in \(C_2\), and so on. If there are multiple objects in set \(C_j\) these items
are tied in the ranking. To accommodate such ties, \textbf{PlackettLuce} models
each choice in the ranking with the Davidson-Luce model as developed recently
by \citet{Firth2019}.

For a set \(S\), let
\[f(S) = \delta_{|S|} \left(\prod_{i \in S} \alpha_i \right)^\frac{1}{|S|},\]
where \(|S|\) is the cardinality of the set \(S\), \(\delta_n\) is a parameter
representing the prevalence of ties of order \(n\) (with \(\delta_1 \equiv 1\)),
and \(\alpha_i\) is the worth of item \(i\). Then, the probability of a
ranking \(R\) with ties up to order \(D\) is given by
\begin{equation}
\prod_{j = 1}^J \frac{f(C_j)}{
\sum_{k = 1}^{\min(D_j, D)} \sum_{S \in {\binom{A_j}{k}}} f(S)}.
\label{eq:PL}
\end{equation}
Here \(D_j\) is the cardinality of \(A_j\), which is the set of items from
which \(C_j\) is chosen, and \(\binom{A_j}{k}\) denotes the set of all
possible choices of \(k\) items from \(A_j\). The value of \(D\) can be set to the
maximum number of tied items observed in the data, so that \(\delta_n = 0\) for
\(n > D\).

The scale of the worth parameters needs to be constrained for the parameters
to be identifiable. \textbf{PlackettLuce} constrains the worth parameters to sum to
one, so that they represent the probability that the corresponding item comes
first in a ranking of all items, given that first place is not tied.

The 2-way tie parameter \(\delta_2\) is related to the probability
that two items \emph{of equal worth} tie for first place, given that the
first place is not a 3-way or higher tie. Specifically, that probability is
\(\delta_2/(2 + \delta_2)\). The 3-way and higher tie parameters are
similarly interpretable, in terms of tie probabilities among equal-worth items.

Further description of this general model for handling multi-item ties,
including its origins and its most important properties, can be found in
\citet{Firth2019}.

\hypertarget{pudding-example}{%
\subsubsection{Pudding example}\label{pudding-example}}

When each ranking contains only two items, the model in Equation
\eqref{eq:PL} reduces to the extended Bradley-Terry model proposed by
\citet{Davidson1970}. In Davidson (\citeyearpar{Davidson1970}, Example 2), a case study is presented
where respondents were asked to test two brands of chocolate pudding from a
total of six brands. The results are provided as the \texttt{pudding} data set in
\textbf{PlackettLuce}. For each pair of brands \(i\) and \(j\), the data set
gives the total number of paired comparisons (\(r_{ij}\)), along with
the frequencies that brand \(i\) was preferred (\(w_{ij}\)), that brand \(j\)
was preferred (\(w_{ji}\)) and that the brands were tied (\(t_{ij}\)).

\begin{verbatim}
R> library(PlackettLuce)
R> head(pudding)
\end{verbatim}

\begin{verbatim}
  i j r_ij w_ij w_ji t_ij
1 1 2   57   19   22   16
2 1 3   47   16   19   12
3 2 3   48   19   19   10
4 1 4   54   18   23   13
5 2 4   51   23   19    9
6 3 4   54   19   20   15
\end{verbatim}

\texttt{PlackettLuce()}, the model-fitting function in \textbf{PlackettLuce},
requires each ranking to be specified by the numeric rank (1, 2, 3, \(\ldots\))
assigned to each item. For the pudding data, it is easier to specify the
ordering (winner, loser) initially. In particular, we can define the winner and
loser for the three sets of paired comparions: the wins for brand \(i\), the wins
for brand \(j\) and the ties:

\begin{verbatim}
R> i_wins <- data.frame(Winner = pudding$i, Loser = pudding$j)
R> j_wins <- data.frame(Winner = pudding$j, Loser = pudding$i)
R> ties <- data.frame(Winner = asplit(pudding[c("i", "j")], 1),
+                     Loser = rep(NA, 15))
R> head(ties, 2)
\end{verbatim}

\begin{verbatim}
  Winner Loser
1   1, 2    NA
2   1, 3    NA
\end{verbatim}

In the last case, the base \texttt{R} function \texttt{asplit()} is used to split the \texttt{i} and
\texttt{j} columns of \texttt{pudding} by row, giving a vector of items that we can specify
as the winner, while the loser is missing.

The combined orderings can be converted to numeric rankings using the
\texttt{as.rankings()} function from \textbf{PlackettLuce} :

\begin{verbatim}
R> pudding_rankings <- as.rankings(rbind(i_wins, j_wins, ties),
+                                  input = "orderings")
R> head(pudding_rankings)
\end{verbatim}

\begin{verbatim}
[1] "1 > 2" "1 > 3" "2 > 3" "1 > 4" "2 > 4" "3 > 4"
\end{verbatim}

\begin{verbatim}
R> tail(pudding_rankings)
\end{verbatim}

\begin{verbatim}
[1] "4 = 5" "1 = 6" "2 = 6" "3 = 6" "4 = 6" "5 = 6"
\end{verbatim}

The result is an object of class \texttt{"rankings"}. The print method displays
the rankings in a readable form, indicating the ordering of the brands.
However, the underlying data structure stores the rankings in the form
of a matrix:

\begin{verbatim}
R> head(unclass(pudding_rankings), 2)
\end{verbatim}

\begin{verbatim}
     1 2 3 4 5 6
[1,] 1 2 0 0 0 0
[2,] 1 0 2 0 0 0
\end{verbatim}

The six columns represent the pudding brands. In each row, \texttt{0} represents an
unranked brand (not in the comparison), \texttt{1} represents the brand(s) ranked
in first place and \texttt{2} represents the brand in second place, if applicable.
\texttt{PlackettLuce()} will also accept rankings in this matrix form, which it will
convert to a rankings object using \texttt{as.rankings()} with \texttt{input\ =\ "rankings"}.

To specify the full set of rankings, we need the frequency of each ranking,
which will be specified to the model-fitting function as a weight vector:

\begin{verbatim}
R> w <- unlist(pudding[c("w_ij", "w_ji", "t_ij")])
\end{verbatim}

Now we can fit the model with \texttt{PlackettLuce()}, passing the rankings object
and the weight vector as arguments. Setting \texttt{npseudo\ =\ 0} means that standard
maximum likelihood estimation is performed and \texttt{maxit\ =\ 7} limits the number of
iterations to obtain the same worth parameters as \citet{Davidson1970}:

\begin{verbatim}
R> pudding_model <- PlackettLuce(pudding_rankings, 
+                                weights = w, npseudo = 0, maxit = 7)
\end{verbatim}

\begin{verbatim}
Warning in PlackettLuce(pudding_rankings, weights = w, npseudo = 0,
maxit = 7): Iterations have not converged.
\end{verbatim}

\begin{verbatim}
R> coef(pudding_model, log = FALSE)
\end{verbatim}

\begin{verbatim}
        1         2         3         4         5         6      tie2 
0.1388005 0.1729985 0.1617420 0.1653930 0.1586805 0.2023855 0.7468147 
\end{verbatim}

Here we have specified \texttt{log\ =\ FALSE} to report the estimates
in the parameterization of Equation~\eqref{eq:PL}, with the worth parameters
constrained to sum to one. In Section~\ref{inference} we discuss why it is
more appropriate to use the log scale for inference.

\hypertarget{choice-of-algorithm}{%
\subsubsection{Choice of algorithm}\label{choice-of-algorithm}}

As noted in Section~\ref{specialized-software}, \texttt{PlackettLuce()}
offers a selection of three fitting algorithms, which can be specified by setting the
argument \texttt{method} to \texttt{"iterative\ scaling"}, \texttt{"BFGS"} or \texttt{"L-BFGS"}.
Iterative scaling is the default because it is a reliable, albeit slow,
method. Convergence is determined by direct comparison of the
observed and expected values of the sufficient statistics for the parameters.

The sufficient statistics are: for each item, the number of outright wins
plus one \(n\)th of the number of \(n\)-way ties (\(n = 1, \ldots, D\));
and for each tie of order \(n\), the number of \(n\)-way ties.
Assessing convergence by how close the sufficient statistics are to their
expected value gives more control over the precision of the parameter
estimates, compared to specifying a tolerance on the change in log-likelihood.

Iterative scaling is slow because it does not use derivatives of
the log-likelihood. However, \texttt{PlackettLuce()} uses Steffensen's method
\citep[see for example][]{Nievergelt1991}
to accelerate the algorithm when it gets close to the final solution.

The BFGS and L-BFGS algorithms are implemented in \textbf{PlackettLuce} via the
workhorse functions \texttt{optim()} (from base \texttt{R}) or \texttt{lbfgs()} (from \textbf{lfbgs}).
Both algorithms use derivatives of the log-likelihood, therefore typically
require fewer iterations than iterative scaling. However, BFGS approximates
the Hessian matrix (matrix of second derivatives), which is
computationally expensive, while L-BFGS uses past estimates of the first
derivatives to avoid evaluating the Hessian. Overall then, L-BFGS typically
has the quickest run time, followed by iterative scaling, then BFGS.

\hypertarget{inference-1}{%
\subsection{Inference}\label{inference-1}}

\hypertarget{standard-errors-and-z-tests}{%
\subsubsection{Standard errors and Z tests}\label{standard-errors-and-z-tests}}

A standard way to report model parameter estimates is to report them along with their
corresponding standard error. This is an indication of the estimate's precision.
However, implicitly this invites comparison with zero. Such comparison is made
explicit in many summary methods for models in \texttt{R}, with the addition of
partial t or Z tests. These are tests of the null hypothesis that a parameter is
equal to zero, provided that the other parameters are in the model. However, this hypothesis
is generally not of interest for the worth parameters in a Plackett-Luce model:
we expect most items to have \emph{some} worth, the question is whether the items
differ in worth. Besides, when the worth parameters are constrained to sum to
one, this constrains each parameter to lie between zero and one. A Z test
based on asymptotic normality of the MLE is not
appropriate for worth parameters near the limits, since it does not account for
the constrained parameter space.

On the log scale however, before identifiability constraints are applied,
there are no bounds on the parameters. We can constrain the worth parameters
by computing simple contrasts with a reference value. By default, the summary
method for \texttt{"PlackettLuce"} objects contrasts with the worth of the first item
(the first element of \texttt{colnames(R)}), which is equivalent to setting the worth
parameter for that item to zero:

\begin{verbatim}
R> summary(pudding_model)
\end{verbatim}

\begin{verbatim}
Call: PlackettLuce(rankings = pudding_rankings, npseudo = 0, 
    weights = w, maxit = 7)

Coefficients:
     Estimate Std. Error z value Pr(>|z|)
1      0.0000         NA      NA       NA
2      0.2202     0.1872   1.176 0.239429
3      0.1530     0.1935   0.790 0.429271
4      0.1753     0.1882   0.931 0.351683
5      0.1339     0.1927   0.695 0.487298
6      0.3771     0.1924   1.960 0.049983
tie2  -0.2919     0.0825  -3.539 0.000402

Residual deviance:  1619.4 on 1484 degrees of freedom
AIC:  1631.4 
Number of iterations: 7
\end{verbatim}

The Z tests for the item parameters test the null hypothesis that the difference
from the worth of item 1 is zero. None of the Z tests provides significant evidence
against this hypothesis, which is consistent with the test for equal preferences
presented in \citet{Davidson1970}. The tie parameter is also shown on the
log scale here, but it is an integral part of the model rather than a parameter
of interest for inference, and its scale is not as relevant as that of the
worth parameters.

The reference value for the item parameters can be changed via the \texttt{ref}
argument, for example, setting to \texttt{NULL} sets the mean worth as the reference:

\begin{verbatim}
R> summary(pudding_model, ref = NULL)
\end{verbatim}

\begin{verbatim}
Call: PlackettLuce(rankings = pudding_rankings, npseudo = 0, 
    weights = w, maxit = 7)

Coefficients:
      Estimate Std. Error z value Pr(>|z|)
1    -0.176581   0.121949  -1.448 0.147619
2     0.043664   0.121818   0.358 0.720019
3    -0.023617   0.126823  -0.186 0.852274
4    -0.001295   0.122003  -0.011 0.991533
5    -0.042726   0.127054  -0.336 0.736657
6     0.200555   0.126594   1.584 0.113140
tie2 -0.291938   0.082499  -3.539 0.000402

Residual deviance:  1619.4 on 1484 degrees of freedom
AIC:  1631.4 
Number of iterations: 7
\end{verbatim}

\hypertarget{quasi-standard-errors}{%
\subsubsection{Quasi standard errors}\label{quasi-standard-errors}}

As shown by the two summaries of the model fitted to the pudding data in
Section~\ref{standard-errors-and-z-tests}, the
standard error of the item parameters changes with the reference level.
In some cases, there will be a natural
reference, for example, when comparing an own brand product with competing
brands. When this is not the case, inference can depend on an arbitrary choice.
This problem can be handled through the use of
\emph{quasi standard errors} \citetext{\citealp{FirthMenezes2004}; \citealp[see also][]{Firth2003}}
that remain constant for a given item regardless of the
reference. In particular, the quasi standard error is not changed by setting an
item as the reference, so even in cases where there is a natural reference, the
uncertainty around the worth of that item can be represented.

Quasi standard errors for the item parameters are implemented via a method for
the \texttt{qvcalc} function from the \textbf{qvcalc} package:

\begin{verbatim}
R> pudding_qv <- qvcalc(pudding_model)
R> summary(pudding_qv)
\end{verbatim}

\begin{verbatim}
Model call: PlackettLuce(rankings = pudding_rankings, npseudo = 0, weights = w, maxit = 7) 
       estimate        SE   quasiSE   quasiVar
    1 0.0000000 0.0000000 0.1328950 0.01766108
    2 0.2202447 0.1872168 0.1327373 0.01761919
    3 0.1529644 0.1935181 0.1395740 0.01948091
    4 0.1752864 0.1882110 0.1330240 0.01769538
    5 0.1338550 0.1927043 0.1399253 0.01957908
    6 0.3771362 0.1924059 0.1392047 0.01937796
Worst relative errors in SEs of simple contrasts (%):  -0.8 0.8 
Worst relative errors over *all* contrasts (%):  -1.7 1.7 
\end{verbatim}

Again by default, the first item is taken as the reference, but this may be
changed via the \texttt{ref} argument. The plot method for the returned object visualizes
the item parameters (log worth parameters) along with comparison intervals:

\begin{verbatim}
R> plot(pudding_qv, xlab = "Brand of pudding", ylab = "Worth (log)",
+      main = NULL)
\end{verbatim}

\begin{figure}

{\centering \includegraphics[width=0.8\textwidth]{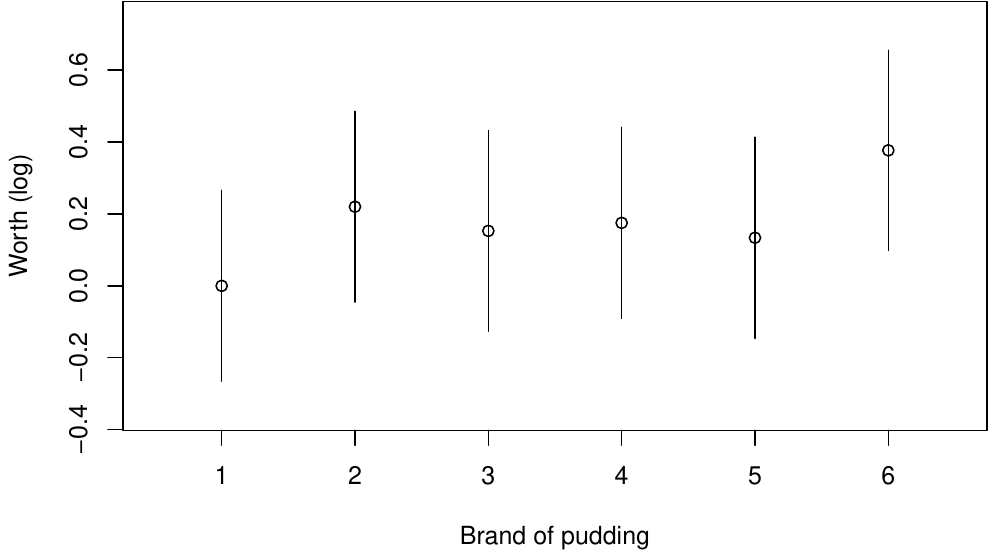} 

}

\caption{Worth of brands of chocolate pudding. Comparison intervals based on quasi standard errors.}\label{fig:pudding-qv}
\end{figure}

The output is shown in Fig. \ref{fig:pudding-qv}. There is evidence of a
difference between any two item parameters if there is no overlap between the
corresponding comparison intervals. The
quasi-variances allow comparisons that are \emph{approximately} correct, for every
possible contrast among the parameters. The routine error report in the last
two lines printed by \texttt{summary(pudding\_qv)} tells us that, in this example, the
approximation error has been very small: the approximation error for the
standard error of any simple contrast among the parameters is less than 0.8\%.

\hypertarget{disconnected-or-weakly-connected-networks-1}{%
\subsection{Disconnected or weakly connected networks}\label{disconnected-or-weakly-connected-networks-1}}

Given a ranking of items, we can infer the wins and losses between pairs
of items in the ranking. For example, the ranking \{item 1\} \(\succ\)
\{item 3, item 4\} \(\succ\) \{item 2\}, implies that item 1 wins against
items 2, 3 and 4; and items 3 and 4 win against item 2. These wins and losses
can be represented as a directed network.

For example, consider the following artificial set of paired comparisons:

\begin{verbatim}
R> ABCD_matrix <- matrix(c(1, 2, 0, 0,    ## A beats B
+                          2, 0, 1, 0,    ## C beats A
+                          1, 0, 0, 2,    ## A beats D
+                          2, 1, 0, 0,    ## B beats A
+                          0, 1, 2, 0),   ## B beats C
+                        byrow = TRUE, ncol = 4,
+                        dimnames = list(NULL, LETTERS[1:4]))
R> ABCD_rankings <- as.rankings(ABCD_matrix)
\end{verbatim}

Here we have defined the paired comparisons as a matrix of
numeric rankings, then used \texttt{as.rankings()} to create a formal rankings object
(\texttt{input} is set to \texttt{"rankings"} by default). The \texttt{as.rankings()} function
ensures that the rankings are coded as dense rankings, i.e.~consecutive
integers with no rank skipped for tied items; sets any rankings with less than two
items to \texttt{NA} since these are uninformative, and adds item names if necessary.

The \texttt{adjacency()} function from \textbf{PlackettLuce} can be used to convert these
rankings to an adjacency matrix, where element \((i, j)\) is the number of times
item \(i\) is ranked higher than item \(j\):

\begin{verbatim}
R> ABCD_adjacency <- adjacency(ABCD_rankings)
R> ABCD_adjacency
\end{verbatim}

\begin{verbatim}
  A B C D
A 0 1 0 1
B 1 0 1 0
C 1 0 0 0
D 0 0 0 0
attr(,"class")
[1] "adjacency" "matrix"   
\end{verbatim}

Using functions from the \textbf{igraph} \texttt{R} package \citep{Csardi2006} we can visualise
the corresponding network (output shown in Fig. \ref{fig:always-loses}):

\begin{verbatim}
R> library(igraph)
R> net <- graph_from_adjacency_matrix(ABCD_adjacency)
R> plot(net, edge.arrow.size = 0.5, vertex.size = 30)
\end{verbatim}

\begin{figure}
\centering
\includegraphics{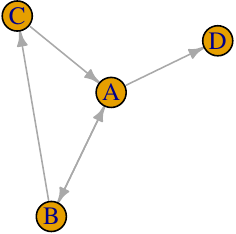}
\caption{\label{fig:always-loses}Network representation of toy rankings.}
\end{figure}

A sufficient condition for the worth parameters (on the log scale) to have finite
MLEs and standard errors is that the network is
strongly connected \citep{Hunter2004}. This means there is a path of wins and a
path of losses between each pair of items. This condition is also necessary
when there are no ties in the data.

In our example data, A, B and C are strongly connected. For example,
C directly loses against B; C never directly beats B, but C does beat A and A,
in turn, beats B, so C indirectly beats B. Similar paths of wins and
losses can be found for all pairs of A, B and C. On the other hand, D is only
observed to lose, therefore the MLE of its log worth would be \(-\infty\), with
infinite standard error.

If one item always wins, the MLE of its log worth would be \(+\infty\) with infinite
standard error. If the items form strongly connected clusters, that are not
connected to each other, or connected only by wins or only by losses
(weakly connected), then the maximum likelihood solution
cannot be represented simply through a vector of relative worths. A variant of
our artificial example illustrates this last point, through a pair of clusters
that are not connected to one another:

\begin{verbatim}
R> disconnected_matrix <- matrix(c(1, 2, 0, 0,    ## A beats B
+                                  2, 1, 0, 0,    ## B beats A
+                                  0, 0, 1, 2,    ## C beats D
+                                  0, 0, 2, 1),   ## D beats C
+                                byrow = TRUE, ncol = 4,
+                                dimnames = list(NULL, LETTERS[1:4]))
R> print(disconnected_rankings <- as.rankings(disconnected_matrix))
\end{verbatim}

\begin{verbatim}
[1] "A > B" "B > A" "C > D" "D > C"
\end{verbatim}

In this example, there is no information that allows the relative worths of
items \(A\) and \(C\) to be estimated. While the within-cluster
worth ratios \(\alpha_A/\alpha_B\) and \(\alpha_C/\alpha_D\) both have MLEs equal to 1 here,
between-cluster ratios such as \(\alpha_A/\alpha_C\) are not estimable.
When \texttt{PlackettLuce()} is applied to such a set of rankings, with argument
\texttt{npseudo\ =\ 0} to specify
the use of maximum likelihood estimation, by design the result is an error message:

\begin{verbatim}
R> try(PlackettLuce(disconnected_rankings, npseudo = 0))
\end{verbatim}

\begin{verbatim}
Error in PlackettLuce(disconnected_rankings, npseudo = 0) : 
  Network is not fully connected - cannot estimate all item parameters with npseudo = 0
\end{verbatim}

The case of weakly connected clusters is similar. If, for example, a single
further ranking `\(A\) beats \(C\)' is added to the example just shown, then the
two clusters become weakly connected. The maximum likelihood solution in that
case has \(\hat\alpha_C/\hat\alpha_A = \hat\alpha_C/\hat\alpha_B = \hat\alpha_D/\hat\alpha_A = \hat\alpha_D/\hat\alpha_B = 0\), and requires the
notion of a `direction of recession' \citep{Geyer2009} to describe it fully.
This is beyond the scope of \texttt{PlackettLuce()}, which again by design gives an
informative error message in such situations.

The remainder of this section describes two distinct approaches to the use
of \texttt{PlackettLuce} with disconnected networks: first, the separate analysis
by maximum likelihood of an identified sub-network that is
strongly connected within itself; and
second, a general approach that uses pseudo-rankings to fully connect
the network of items being compared.

We return to the small `toy' set of rankings depicted in Figure
\ref{fig:always-loses}. The \texttt{connectivity()} function from \textbf{PlackettLuce}
checks how connected the items are by the wins and losses recorded in the
adjacency matrix:

\begin{verbatim}
R> connectivity(ABCD_adjacency)
\end{verbatim}

\begin{verbatim}
Network of items is not strongly connected
\end{verbatim}

\begin{verbatim}
$membership
A B C D 
1 1 1 2 

$csize
[1] 3 1

$no
[1] 2
\end{verbatim}

If the network of wins and losses is not strongly connected, information on the
clusters within the network is returned. In this example, we might proceed by
excluding item D:

\begin{verbatim}
R> ABC_rankings <- ABCD_rankings[, -4]
\end{verbatim}

\begin{verbatim}
Rankings with only 1 item set to `NA`
\end{verbatim}

\begin{verbatim}
R> ABC_rankings 
\end{verbatim}

\begin{verbatim}
[1] "A > B" "C > A" NA      "B > A" "B > C"
\end{verbatim}

Since \texttt{ABCD\_rankings} is a rankings object, the subset method checks the
rankings are still valid after excluding item D. In this case,
the paired comparison with item A only has one item left in the ranking, so it
is set to \texttt{NA}.

The handling of any rankings that are \texttt{NA} is determined by the
\texttt{na.action} argument of \texttt{PlackettLuce()}. This is set to \texttt{"na.omit"} by
default, so that missing rankings are dropped. In our example, the network
corresponding to the remaining rankings is
strongly connected, so the MLEs in the Plackett-Luce model all have finite
worth estimates and standard errors:

\begin{verbatim}
R> ABC_model <- PlackettLuce(ABC_rankings , npseudo = 0)
R> summary(ABC_model)
\end{verbatim}

\begin{verbatim}
Call: PlackettLuce(rankings = ABC_rankings, npseudo = 0)

Coefficients:
  Estimate Std. Error z value Pr(>|z|)
A   0.0000         NA      NA       NA
B   0.8392     1.3596   0.617    0.537
C   0.4196     1.5973   0.263    0.793

Residual deviance:  5.1356 on 2 degrees of freedom
AIC:  9.1356 
Number of iterations: 3
\end{verbatim}

However, it is not necessary to exclude item D, since \textbf{PlackettLuce} provides
a way to handle disconnected/weakly connected networks through the addition of
pseudo-rankings. This works by adding wins and losses between each item and a
hypothetical or ghost item with fixed worth, making the network strongly
connected. This method has an interpretation as a Bayesian prior, in
particular, an exchangeable prior in which all items have equal worth.

The \texttt{npseudo} argument defines the number of wins and loses with the
ghost item that are added for each real item.
The larger \texttt{npseudo} is, the stronger the influence
of the prior; by default \texttt{npseudo} is set to 0.5, so that each
pseudo-ranking is weighted by 0.5. This fractional weight is enough to connect
the network. Note that this weight is only applied to the pseudo-rankings, the
\texttt{weights} argument defines the weights (frequencies) for the observed rankings.

In this toy example, when we fit the model to the full data using
pseudo-rankings the parameters for items B and C change quite considerably
compared to the previous fit:

\begin{verbatim}
R> ABCD_model <- PlackettLuce(ABCD_rankings)
R> coef(ABCD_model)
\end{verbatim}

\begin{verbatim}
         A          B          C          D 
 0.0000000  0.5184185  0.1354707 -1.1537565 
\end{verbatim}

This is because there are only 5 rankings, so weighting each pseudo-ranking by
0.5 gives them relatively large weight. In more realistic examples, with many
rankings, the default prior will have a weak shrinkage effect,
shrinking the items' worth parameters towards \(1/J\),
where \(J\) is the number of items.

Although it is only necessary to use pseudo-rankings when the network is
not strongly connected, the default behaviour is always to use them
(with a weight of 0.5).
This is because the shrinkage that the pseudo-rankings deliver
typically reduces both the bias and the variance of the
estimators of the worth parameters.

\hypertarget{nascar-example}{%
\subsubsection{NASCAR example}\label{nascar-example}}

For a practical example of modelling rankings where the underlying network
is weakly connected, we consider the NASCAR data from \citet{Hunter2004}. The data
are results of the 36 races in the 2002 NASCAR season in the United
States. Each race involves 43 drivers out of a total of 87 drivers. The
\texttt{nascar} data provided by \textbf{PlackettLuce} records the results as an ordering
of the drivers in each race:

\begin{verbatim}
R> data(nascar)
R> ncol(nascar)
\end{verbatim}

\begin{verbatim}
[1] 43
\end{verbatim}

\begin{verbatim}
R> nascar[1:2, 1:9]
\end{verbatim}

\begin{verbatim}
     rank1 rank2 rank3 rank4 rank5 rank6 rank7 rank8 rank9
[1,]    83    18    20    48    53    51    67    72    32
[2,]    52    72     4    82    60    31    32    66     3
\end{verbatim}

For example, in the first race, driver 83 came first, followed by driver 18 and
so on. The names corresponding to the driver IDs are available as an attribute of
\texttt{nascar}. We can provide these names when converting the orderings to rankings
via the \texttt{items} argument:

\begin{verbatim}
R> nascar_rankings <- as.rankings(nascar, input = "orderings", 
+                                 items = attr(nascar, "drivers"))
R> nascar_rankings[1:2]
\end{verbatim}

\begin{verbatim}
[1] "Ward Burton > Elliott Sadler > Geoff ..."
[2] "Matt Kenseth > Sterling Marlin > Bob ..."
\end{verbatim}

Maximum likelihood estimation cannot be used in this example, because four
drivers placed last in each race they entered. For this reason, \citet{Hunter2004} dropped these
four drivers to fit the Plackett-Luce model, which we can reproduce as follows:

\begin{verbatim}
R> driver_1_to_83_rankings <- nascar_rankings[, 1:83]
R> driver_1_to_83_model <- PlackettLuce(driver_1_to_83_rankings, npseudo = 0)
\end{verbatim}

In order to demonstrate the correspondence with the results from \citet{Hunter2004}, we
order the item parameters by the driver's average rank:

\begin{verbatim}
R> average_rank <- apply(nascar_rankings, 2, function(x) mean(x[x > 0]))
R> id <- order(average_rank)
R> ordered_drivers <- names(average_rank)[id]
R> driver_1_to_83_coef <- round(coef(driver_1_to_83_model), 2)
R> driver_1_to_83_coef[ordered_drivers[1:3]]
\end{verbatim}

\begin{verbatim}
    PJ Jones Scott Pruett  Mark Martin 
        4.15         3.62         2.08 
\end{verbatim}

\begin{verbatim}
R> driver_1_to_83_coef[ordered_drivers[81:83]]
\end{verbatim}

\begin{verbatim}
 Dave Marcis Dick Trickle    Joe Varde 
        0.03        -0.31        -0.15 
\end{verbatim}

Now we fit the Plackett-Luce model to the full data, using the default
pseudo-rankings method.

\begin{verbatim}
R> nascar_model <- PlackettLuce(nascar_rankings)
\end{verbatim}

For items that were in the previous model, we see that the log worth parameters
generally shrink towards zero:

\begin{verbatim}
R> nascar_coef <- round(coef(nascar_model), 2)
R> nascar_coef[ordered_drivers[1:3]]
\end{verbatim}

\begin{verbatim}
    PJ Jones Scott Pruett  Mark Martin 
        3.20         2.77         1.91 
\end{verbatim}

\begin{verbatim}
R> nascar_coef[ordered_drivers[81:83]]
\end{verbatim}

\begin{verbatim}
 Dave Marcis Dick Trickle    Joe Varde 
        0.02        -0.38        -0.12 
\end{verbatim}

The new items have relatively large negative log worth, but the estimates are
nonetheless finite with finite standard error:

\begin{verbatim}
R> coef(summary(nascar_model))[84:87,]
\end{verbatim}

\begin{verbatim}
                 Estimate Std. Error    z value  Pr(>|z|)
Andy Hillenburg -2.171065   1.812994 -1.1975028 0.2311106
Gary Bradberry  -1.744754   1.855365 -0.9403828 0.3470212
Jason Hedlesky  -1.590764   1.881708 -0.8453828 0.3978972
Randy Renfrow   -1.768629   1.904871 -0.9284767 0.3531604
\end{verbatim}

\begin{figure}
\includegraphics[width=\textwidth]{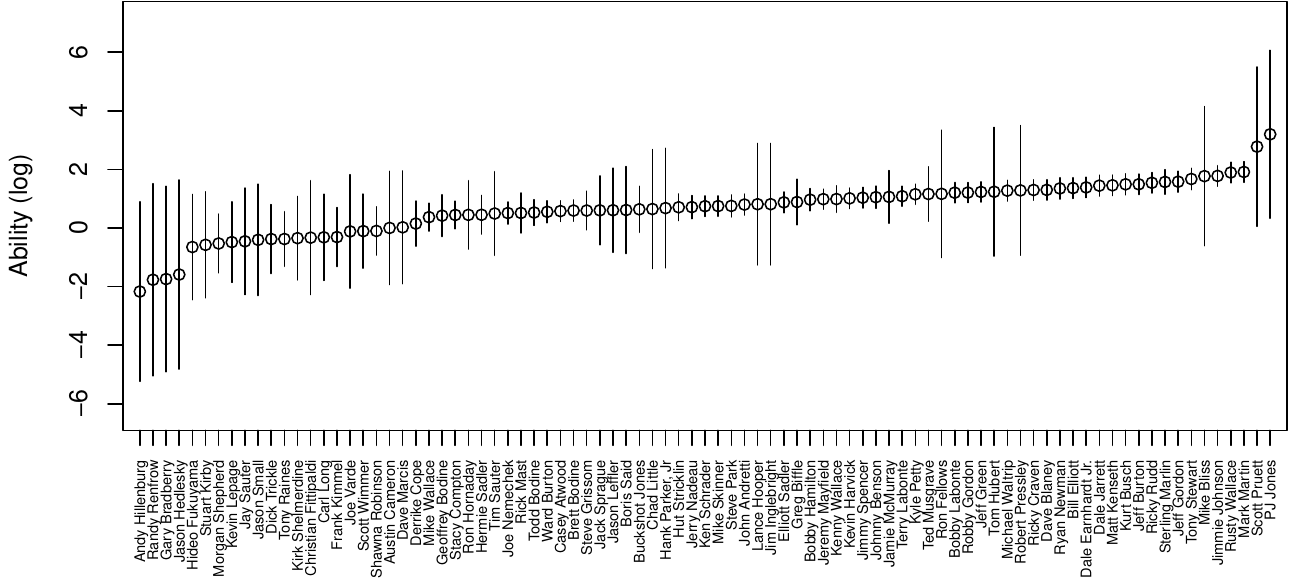} \caption{Ability of drivers based on NASCAR 2002 season. Intervals based on quasi standard errors.}\label{fig:nascar-qv}
\end{figure}

We can plot the quasi-variances to visualise the relative ability of all
drivers (output shown in Fig. \ref{fig:nascar-qv}):

\begin{verbatim}
R> nascar_qv <- qvcalc(nascar_model)
R> nascar_qv$qvframe <- nascar_qv$qvframe[order(coef(nascar_model)),]
R> plot(nascar_qv, xlab = NULL, ylab = "Ability (log)", main = NULL,
+       xaxt = "n", xlim = c(3, 85))
R> axis(1, at = seq_len(87), labels = rownames(nascar_qv$qvframe),
+       las = 2, cex.axis = 0.6)
\end{verbatim}

As in the pudding example, we can use \texttt{summary(nascar\_qv)} to see a
report on the accuracy of the quasi-variance approximation. In this example the error of
approximation, across the standard errors of all 3741 possible simple
contrasts (contrasts between pairs of the 87 driver-specific parameters),
ranges between -0.7\% and +6.7\%. This is still remarkably accurate and
means that the plot of comparison intervals is a good visual guide to the
uncertainty about drivers' relative abilities. The results of using
\texttt{summary(nascar\_qv)} are not shown here, as they would occupy too much space.

\hypertarget{plackett-luce-trees}{%
\section{Plackett-Luce Trees}\label{plackett-luce-trees}}

A Plackett-Luce model that assumes the same worth parameters across all rankings
may sometimes be an over-simplification. For example, if rankings are made by
different judges, the worth parameters may vary between groups of judges with
different characteristics. Alternatively, the worth parameters may vary by
conditions in which the ranking took place, such as the weather at the time of a
race.

Model-based partitioning provides an automatic way to determine subgroups
with significantly different sets of worth parameters. The rankings are
recursively split into two groups according to their value on covariates
whose values vary between rankings, and a Plackett-Luce model is fitted to
each subgroup formed. The method proceeds as follows:

\begin{enumerate}
\def\labelenumi{\arabic{enumi}.}
\item
  Fit a Plackett-Luce model to the full data.
\item
  Assess the stability of the worth parameters with respect to each
  covariate. That is, test for a structural change in the ranking-level
  contributions to the first derivatives of the log-likelihood, when these
  contributions are ordered by covariate values.
\item
  If there is significant instability, split the full data by the covariate
  with the strongest instability, using the cut-point that gives the highest
  improvement in model fit.
\item
  Repeat steps 1-3 for each subgroup until there are no more significant
  instabilities, or a split produces a subgroup below a given size threshold.
\end{enumerate}

The result is a Plackett-Luce tree, since the recursive binary splits may be
represented in a tree diagram and a Plackett-Luce model is fitted to the
subgroups defined by each branch. This is an extension of Bradley-Terry trees,
implemented in the \texttt{R} package \textbf{psychotree} \citep{Zeileis2018} and described in
more detail by \citet{Strobl2011}.

\hypertarget{beans-example}{%
\subsection{Beans example}\label{beans-example}}

To illustrate Plackett-Luce trees, we consider data from a trial in Nicaragua
of ten different varieties of bean, run by Bioversity International \citep{vanEtten2016}.
Farmers were asked to grow three out of the ten trial varieties in one of the growing
seasons, Primera (May - August), Postrera (September - October) or Apante
(November - January). At the end of the season, they were asked which variety
they thought was best and which variety they thought was worst, to give a
ranking of the three varieties. In addition, they were asked to compare each
trial variety to the standard local variety and say whether it was better or
worse.

The data are provided as the dataset \texttt{beans} in \textbf{PlackettLuce}. The data
require some preparation to collate the rankings, which is detailed in
Appendix \ref{app:beans}. The same code is provided in the examples section
of the help file of \texttt{beans}

\begin{verbatim}
R> example("beans", package = "PlackettLuce")
\end{verbatim}

The result is a \texttt{rankings} object \texttt{R} with all rankings of the three
trial varieties and the output of their comparison with the local
variety.

In order to fit a Plackett-Luce tree, we need to create a \texttt{"grouped\_rankings"}
object, that defines how the rankings map to the covariate values. In this case
we wish to group by each record in the original data set. So we group by an
index that identifies the record number for each of the four rankings from each
farmer (one ranking of order three plus three pairwise rankings with the
local variety):

\begin{verbatim}
R> n <- nrow(beans)
R> beans_grouped_rankings <- group(R, index = rep(1:n, times = 4))
R> format(head(beans_grouped_rankings, 2), width = 45)
\end{verbatim}

\begin{verbatim}
                                                                  1 
 "PM2 Don Rey > SJC 730-79 > BRT 103-182, Local > BRT 103-182, ..." 
                                                                  2 
"INTA Centro Sur > INTA Sequia > INTA Rojo, Local > INTA Rojo, ..." 
\end{verbatim}

For each record in the original data, we have three covariates: \texttt{season}
the season-year the beans were planted, \texttt{year} the year of planting,
and \texttt{maxTN} the maximum temperature at night during the vegetative cycle.
The following code fits a Plackett-Luce tree with up to three nodes and at
least 5\% of the records in each node:

\begin{verbatim}
R> beans$year <- factor(beans$year)
R> beans_tree <- pltree(beans_grouped_rankings ~ .,
+                       data = beans[c("season", "year", "maxTN")],
+                       minsize = 0.05*n, maxdepth = 3)
R> beans_tree
\end{verbatim}

\begin{verbatim}
Plackett-Luce tree

Model formula:
G ~ .

Fitted party:
[1] root
|   [2] maxTN <= 18.7175
|   |   [3] season in Pr - 16: n = 47
|   |            ALS 0532-6     BRT 103-182 INTA Centro Sur    INTA Ferroso 
|   |          0.0000000000   -1.1402562056   -0.7356643306   -1.1986768665 
|   |        INTA Matagalpa     INTA Precoz       INTA Rojo     INTA Sequia 
|   |         -0.8016667298    0.1226994334    0.6378654995    0.0003316762 
|   |                 Local     PM2 Don Rey      SJC 730-79 
|   |          0.4788504910   -1.0636844356   -1.2405493189 
|   |   [4] season in Po - 15, Ap - 15, Po - 16, Ap - 16: n = 489
|   |            ALS 0532-6     BRT 103-182 INTA Centro Sur    INTA Ferroso 
|   |            0.00000000      0.12989016      0.21133555      0.08487525 
|   |        INTA Matagalpa     INTA Precoz       INTA Rojo     INTA Sequia 
|   |            0.14435409     -0.16024617      0.11497038      0.50611822 
|   |                 Local     PM2 Don Rey      SJC 730-79 
|   |            0.63702034     -0.11244954      0.15430957 
|   [5] maxTN > 18.7175: n = 306
|            ALS 0532-6     BRT 103-182 INTA Centro Sur    INTA Ferroso 
|            0.00000000      0.38374777      0.43829266      0.01041950 
|        INTA Matagalpa     INTA Precoz       INTA Rojo     INTA Sequia 
|            0.16107300      0.02310481      0.10892957      0.49692201 
|                 Local     PM2 Don Rey      SJC 730-79 
|            0.26238130      0.26594548     -0.07345855 

Number of inner nodes:    2
Number of terminal nodes: 3
Number of parameters per node: 11
Objective function (negative log-likelihood): 3152.027
\end{verbatim}

The algorithm identifies three nodes, with the first split defined by high
night-time temperatures and the second split separating the single Primera season
from the others. So for early planting in regions where the night-time
temperatures were not too high, INTA Rojo (variety 7) was most preferred, closely
followed by the local variety. During the regular growing seasons (Postrera and
Apante) in regions where the night-time temperatures were not too high, the
local variety was most preferred, closely followed by INTA Sequia (8). Finally
in regions where the maximum night-time temperature was high, INTA Sequia (8)
was most preferred, closely followed by BRT 103-182 (2) and INTA Centro Sur
(3). A plot method is provided to visualise the tree (output shown in Fig. \ref{fig:plot-pltree}):

\begin{verbatim}
R> plot(beans_tree, names = FALSE, abbreviate = 2, ylines = 2)
\end{verbatim}

\begin{figure}
\includegraphics[width=\textwidth]{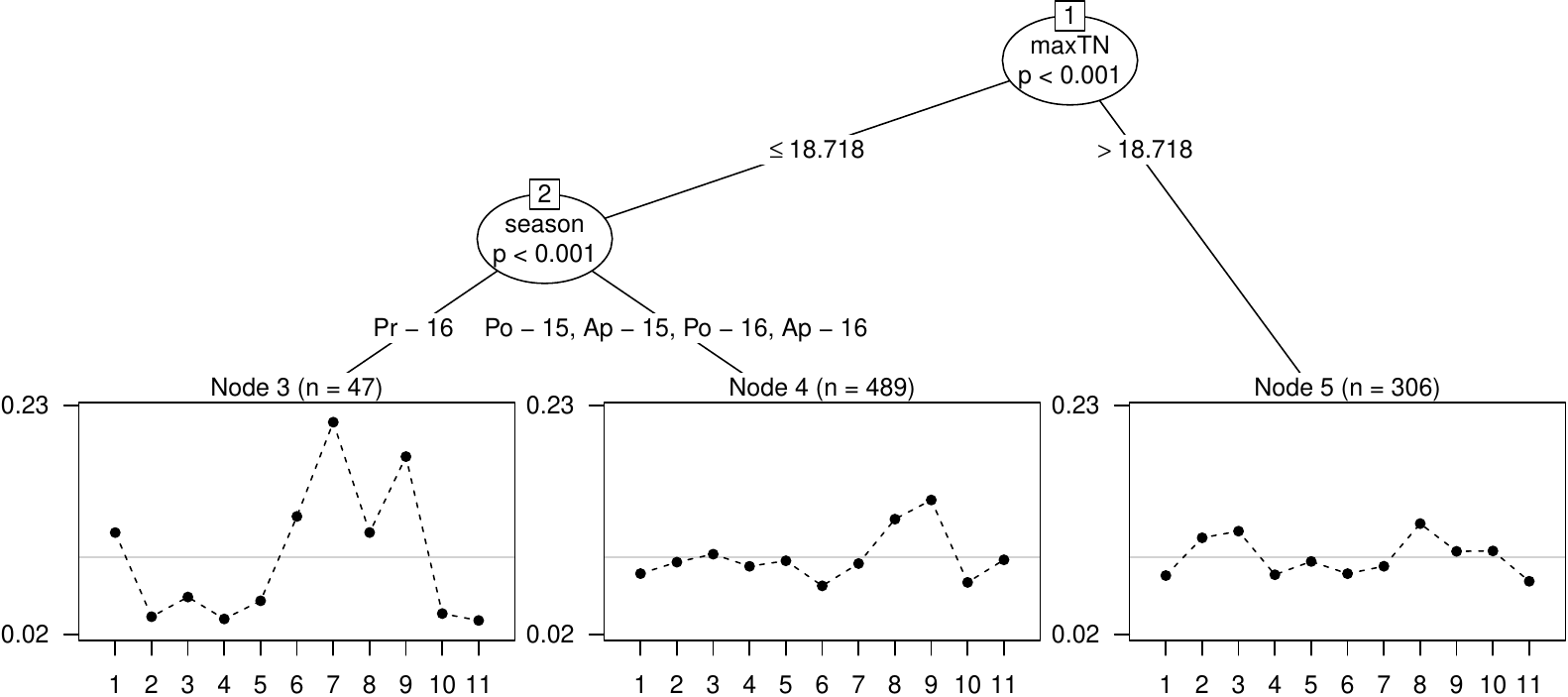} \caption{Worth parameters for the ten trial varieties and the local variety for each node in the Plackett-Luce tree. Varieties are 1: ALS 0532-6, 2: BRT 103-182, 3: INTA Centro Sur, 4: INTA Ferroso, 5: INTA Matagalpa, 6: INTA Precoz, 7: INTA Rojo, 8: INTA Sequia, 9: Local, 10: PM2 Don Rey, 11: SJC 730-79.}\label{fig:plot-pltree}
\end{figure}

Note that when working with high dimensional data in particular, tree models can
be quite unstable. For a more robust model, an ensemble method such as bagging
or random forests could be employed \citep{Strobl2009}.

\hypertarget{discussion}{%
\section{Discussion}\label{discussion}}

\textbf{PlackettLuce} is a feature-rich package for the analysis of ranking data. The
package provides methods for importing and handling partial rankings
data, and for estimation and inference from a generalization of the
Plackett-Luce model that can handle subset rankings and ties of
arbitrary order. Disconnected item networks are handled by appropriately
augmenting the data with pseudo-rankings for a hypothetical item. The package
also facilitates the construction of Plackett-Luce trees, which allow item
worth parameters to vary between rankings, according to available covariate
information on the rankings.

A disadvantage of modelling ties explicitly is that the normalising constant
in Equation~\eqref{eq:PL}, that normalises over all possible choices for a
given rank, involves a number of terms that increases rapidly with the tie
order \(D\). This causes a corresponding increase in the run time of
\texttt{PlackettLuce()}. Also, as noted in Section~\ref{inference}, our current
method for estimating the variance-covariance matrix relies on the technique of
expanding rankings into counts. When this expansion is large,
estimating the standard errors for the model parameters can take almost as
long as fitting the model, or can even become infeasible due to memory limits.
Improving the efficiency of both the model-fitting and the variance-covariance
estimation is a focus of current work.

Meanwhile, we expect that \texttt{PlackettLuce()} and all its methods work well on
data with ties up to order 4, for moderate numbers of items and unique rankings.
As a guide, for 5000 sub-rankings of 10 items, out of 100 items in total, with ties
up to order 4, a test model fit took 18.1s and variance-covariance estimation took
14.8s on the same machine used in the performance evaluation of
Section~\ref{performance-of-software-to-fit-the-plackett-luce-model}. This
makes \textbf{PlackettLuce} suitable for applications such as sport, where ties of
order greater than 3 are typically extremely rare. For data with higher-order
ties, a method that does not explicitly model tied events, such as the
rank-ordered logit model or the Mallows distance-based model, may be more suitable.

Our plans for future development of \textbf{PlackettLuce} include support for
online estimation from streams of partial rankings and formally accounting
for spatio-temporal heterogeneity in worth parameters. Algorithms for
online estimation will be more efficient, enabling the package to handle
larger data sets. To realise this potential,
alternative data structures may be required for partial rankings, which
are currently represented as full rankings with some zero ranks.

Other potential directions for future work include implementing some
of the features offered by other software. In particular, top \(n\)
rankings often arise in practice and are compatible with the algorithms
that are currently implemented. Modelling top \(n\) rankings is more
computationally demanding than modelling subset rankings of the same length,
since each choice in the ranking selects from a wider set of alternatives. So
supporting top \(n\) rankings will have wider practical application if it is
paired with a more efficient algorithm and/or a more efficient
implementation using compiled code.

\hypertarget{appendix}{%
\section{Appendix}\label{appendix}}

\hypertarget{app:run-times}{%
\subsection{Evaluation of run times to fit the Plackett-Luce model}\label{app:run-times}}

Data for the package comparison in Table \ref{tab:timings-kable} was downloaded
from PrefLib \citep{Mattei2013} using the \texttt{read.soc} function provided in
\textbf{PlackettLuce} to read in files with the ``Strict Orders - Complete List''
format.

\begin{verbatim}
R> library(PlackettLuce)
R> # read in example data sets
R> pref <- "http://www.preflib.org/data/election/"
R> netflix <- read.soc(file.path(pref,
+                                "netflix/ED-00004-00000101.soc"))
R> tshirt <- read.soc(file.path(pref,
+                               "shirt/ED-00012-00000001.soc"))
R> sushi <- read.soc(file.path(pref, 
+                              "sushi/ED-00014-00000001.soc"))
\end{verbatim}

A wrapper was defined for each function in the comparison, to prepare the
rankings and run each function with reasonable defaults. The Plackett-Luce model
was fitted to aggregated rankings where possible (for \texttt{PlackettLuce()},
\texttt{hyper2()}, and \texttt{pmr()}). Arguments were set to obtain the MLE,
with the default convergence criteria. The default iterative
scaling algorithm was used for \texttt{PlackettLuce()}.

\begin{verbatim}
R> pl <- function(dat, ...){
+     # convert ordered items to ranking
+     R <- as.rankings(dat[,-1], "ordering")
+     # fit without adding pseudo-rankings, weight by count
+     PlackettLuce(R, npseudo = 0, weights = dat$Freq)
+ }
R> hyper2 <- function(dat, ...){
+     requireNamespace("hyper2")
+     # create likelihood object based on ordered items & counts
+     H <- hyper2::hyper2(pnames = 
+                             paste0("p", seq_len(ncol(dat) - 1)))
+     for (i in seq_len(nrow(dat))){
+         x <-  dat[i, -1][dat[i, -1] > 0]
+         H <- H + hyper2::order_likelihood(x, times = dat[i, 1])
+     }
+     # find parameters to maximise likelihood
+     p <- hyper2::maxp(H)
+     structure(p, loglik = hyper2::loglik(H, p[-length(p)]))
+ }
R> plmix <- function(dat, ...){
+     requireNamespace("PLMIX")
+     # disaggregate data (no functionality for weights or counts)
+     r <- rep(seq_len(nrow(dat)), dat$Freq)
+     # maximum a posteriori estimate, with non-informative prior
+     # K items in each ranking, single component distribution
+     # default init values do not always work; specify as uniform
+     K <- ncol(dat) - 1
+     PLMIX::mapPLMIX(as.matrix(dat[r, -1]), K = K, G = 1,
+                     init = list(p = rep.int(1/K, K)),
+                     plot_objective = FALSE)
+ }
R> pmr <- function(dat, ...){
+     requireNamespace("pmr")
+     # convert ordered items to ranking
+     R <- as.rankings(dat[,-1], "ordering")
+     # create data frame with counts as required by pl
+     X <- as.data.frame(unclass(R))
+     X$n <- dat$Freq
+     capture.output(res <- pmr::pl(X))
+     res
+ }
R> statrank <- function(dat, iter){
+     requireNamespace("StatRank")
+     # disaggregate data (no functionality for weights or counts)
+     r <- rep(seq_len(nrow(dat)), dat$Freq)
+     capture.output(res <- StatRank::Estimation.PL.MLE(
+         as.matrix(dat[r, -1]), iter = iter))
+     res
+ }
R> rologit <- function(dat, ...){
+     requireNamespace("ROlogit")
+     # disaggregate data (no functionality for weights or counts)
+     r <- rep(seq_len(nrow(dat)), dat$Freq)
+     # convert ordered items to ranking
+     R <- as.rankings(dat[r,-1], "ordering")
+     # set up variables for cox PH model and fit
+     nc <- ncol(R)
+     nr <- nrow(R)
+     mf <- data.frame(rank = nc - c(t(R)) + 1,
+                      item = factor(rep(seq_len(nc), nr)),
+                      ranking = factor(rep(seq_len(nr),
+                                           each = nc)))
+     ROlogit::rologit("rank", "item", svar = "ranking", dat = mf,
+                      plot = FALSE)
+ }
R> mm <- function(dat, ...){
+     requireNamespace("mixedMem")
+     # disaggregate data (no support for weights or counts)
+     r <- rep(seq_len(nrow(dat)), dat$Freq)
+     X <- as.matrix(dat[r,-1])
+     nitems <- ncol(X)
+     n <- nrow(X)
+     # Set up parameters for each outcome and repeated measure
+     #  - here just one outcome (ranking) and one measure.
+     # Nijr = number of items in each ranking
+     #  - as complete rankings here always nitems
+     Nijr <- array(nitems, dim = c(n, 1, 1))
+     # inital values for worth parameters
+     theta <- array(rep(1/nitems, nitems),
+                    dim = c(1, 1, nitems))
+     # mixedMemModel assumes items labelled 0 to (nitems -1)
+     X2 <- array(X - 1, dim = c(n, 1, 1, nitems))
+     model <- mixedMem::mixedMemModel(
+         Total = n, # no. observations
+         J = 1, # n of outcomes
+         Rj = 1, # n of repeated measures
+         Nijr= Nijr, # number of items ranked
+         K = 1, # sub-populations, fix to 1
+         Vj = nitems, # total number of items
+         dist = "rank", # type of outcome
+         obs = X2, # ordering of items
+         alpha = 0.5, # hyperparameter for membership
+         theta = theta)
+     # gives warnings, but solution is correct
+     mixedMem::mmVarFit(model)
+ }
\end{verbatim}

Run times were recorded on a Linux machine with a 2.10 GHz Intel Core i7 CPU
and 16 GB RAM. When recording run times, the number of iterations for
\texttt{Estimation.PL.MLE()}, the model-fitting function from \textbf{StatRank},
was set so that the log-likelihood on exit was equal to the log-likelihood
returned by the other functions with relative tolerance 1e-6.

\begin{verbatim}
R> timings <- function(dat, iter = NULL,
+                     fun = c("pl", "hyper2", "plmix", "pmr",
+                             "statrank", "rologit", "mm")){
+     res <- list()
+     for (nm in c("pl", "hyper2", "plmix", "pmr", "statrank",
+                  "rologit", "mm")){
+         if (nm %in% fun){
+             res[[nm]] <- suppressWarnings(
+                 system.time(
+                     do.call(nm, list(dat, iter)))[["elapsed"]])
+         } else res[[nm]] <- NA
+     }
+     res
+ }
R> no_pmr <- c("pl", "hyper2", "plmix", "statrank",
+              "rologit", "mm")
R> netflix_timings <- timings(netflix, 6)
R> # pmr excluded here as takes a long time to fail
R> tshirt_timings <- timings(tshirt, 341, fun = no_pmr)
R> sushi_timings <- timings(sushi, 5, fun = no_pmr)
\end{verbatim}

\hypertarget{app:beans}{%
\subsection{\texorpdfstring{\texttt{beans} data preparation}{beans data preparation}}\label{app:beans}}

First we handle the best and worst rankings. These give the variety the farmer
thought was best or worst, coded as A, B or C for the first, second or third
variety assigned to the farmer, respectively.

\begin{verbatim}
R> data(beans)
R> head(beans[c("best", "worst")], 2)
\end{verbatim}

\begin{verbatim}
  best worst
1    C     A
2    B     A
\end{verbatim}

We fill in the missing item using the \texttt{complete()} function from \emph{PlackettLuce}:

\begin{verbatim}
R> beans$middle <- complete(beans[c("best", "worst")],
+                           items = c("A", "B", "C"))
R> head(beans[c("best", "middle", "worst")], 2)
\end{verbatim}

\begin{verbatim}
  best middle worst
1    C      B     A
2    B      C     A
\end{verbatim}

This gives an ordering of the three varieties the farmer was given. The names of
these varieties are stored in separate columns of \texttt{beans}:

\begin{verbatim}
R> head(beans[c("variety_a", "variety_b", "variety_c")], 2)
\end{verbatim}

\begin{verbatim}
    variety_a       variety_b   variety_c
1 BRT 103-182      SJC 730-79 PM2 Don Rey
2   INTA Rojo INTA Centro Sur INTA Sequia
\end{verbatim}

We can use the \texttt{decode()} function from \emph{PlackettLuce} to decode the orderings,
replacing the coded values with the actual varieties:

\begin{verbatim}
R> order3 <- decode(beans[c("best", "middle", "worst")],
+                   items = beans[c("variety_a", "variety_b", 
+                                   "variety_c")],
+                   code = c("A", "B", "C"))
\end{verbatim}

The pairwise comparisons with the local variety are stored in another set of
columns

\begin{verbatim}
R> head(beans[c("var_a", "var_b", "var_c")], 2)
\end{verbatim}

\begin{verbatim}
  var_a  var_b  var_c
1 Worse  Worse Better
2 Worse Better Better
\end{verbatim}

To convert these data to orderings we first create vectors of the trial variety
and the outcome in each paired comparison:

\begin{verbatim}
R> trial_variety <- unlist(beans[c("variety_a", "variety_b", 
+                                  "variety_c")])
R> outcome <- unlist(beans[c("var_a", "var_b", "var_c")])
\end{verbatim}

We can then derive the winner and loser in each comparison:

\begin{verbatim}
R> order2 <- data.frame(Winner = ifelse(outcome == "Worse",
+                                       "Local", trial_variety),
+                       Loser = ifelse(outcome == "Worse",
+                                      trial_variety, "Local"),
+                       stringsAsFactors = FALSE, row.names = NULL)
R> head(order2, 2)
\end{verbatim}

\begin{verbatim}
  Winner       Loser
1  Local BRT 103-182
2  Local   INTA Rojo
\end{verbatim}

Finally we covert each set of orderings to rankings and combine them:

\begin{verbatim}
R> R <- rbind(as.rankings(order3, input = "orderings"),
+             as.rankings(order2, input = "orderings"))
R> head(R)
\end{verbatim}

\begin{verbatim}
[1] "PM2 Don Rey > SJC 730-79 > BRT 103-182"  
[2] "INTA Centro Sur > INTA Sequia > INTA ..."
[3] "INTA Ferroso > INTA Matagalpa > BRT  ..."
[4] "INTA Rojo > INTA Centro Sur > ALS 0532-6"
[5] "PM2 Don Rey > INTA Sequia > SJC 730-79"  
[6] "ALS 0532-6 > INTA Matagalpa > INTA Rojo" 
\end{verbatim}

\begin{verbatim}
R> tail(R)
\end{verbatim}

\begin{verbatim}
[1] "INTA Sequia > Local"    "INTA Sequia > Local"   
[3] "BRT 103-182 > Local"    "Local > INTA Matagalpa"
[5] "Local > INTA Rojo"      "Local > SJC 730-79"    
\end{verbatim}

\renewcommand\refname{References}
\bibliographystyle{spbasic}
\bibliography{bib-papers,bib-packages}

\end{document}